\documentclass[11pt,a4paper]{article}
\usepackage{a4wide}

\begin{document}

\title{Two Puzzles About Computation}
\author{Samson Abramsky}
\date{}
\maketitle

\section{Introduction}

Turing's classical analysis of computation \cite{turing1937computable} gives a compelling account of the nature of the computational process; of \emph{how} we compute. This allows the notion of \emph{computability}, of what can in principle be computed, to be captured in a mathematically precise fashion.

The purpose of this note is to raise two different questions, which are rarely if ever considered, and to which, it seems, we lack convincing, systematic answers. These questions can be posed as:
\begin{itemize}
\item Why do we compute?
\item What do we compute?
\end{itemize}

The point is not so much that we have no answers to these puzzles, as that we have no established body of theory which gives satisfying, systematic answers, as part of a broader understanding. 
By raising these questions, we hope to stimulate some thinking in this direction.

These puzzles were raised in \cite{abramsky2008information}; see also \cite{adriaanscomputation}.

\section{Why Do We Compute?}
The first puzzle is simply stated:
\begin{center}
\fbox{\textbf{Why do we compute?}}
\end{center}
By this we mean: why do we perform (or build machines and get them to perform) actual, physically embodied computations?

There is, indeed, an obvious answer to this question: 
\begin{center}
To gain information (which, therefore, we did not previously have).
\end{center}
But --- how is this possible?\footnote{Indeed, I was once challenged on this point by an eminent physicist (now knighted), who demanded to know how I could speak of information increasing in computation when Shannon Information theory tells us that it cannot! My failure to answer this point very convincingly at the time led me to continue to ponder the issue, and eventually gave rise to this discussion.}
Two problems seems to  arise, one stemming from physics, and one from logic.
\begin{description}
\item[Problem 1:] Doesn't this contradict the second law of thermodynamics?
\item[Problem 2:] Isn't the output \emph{implied} by the input?
\end{description}
We shall discuss each of these in turn.

\subsection*{Problem 1}
The problem is that, presumably, information is conserved in the \emph{total} system.
The natural response is that, nevertheless, there \emph{can} be information flow between, and information increase in, \emph{subsystems}; just as a body can gain heat from its environment.
More precisely, while the entropy of an isolated (total) system cannot decrease, a sub-system \emph{can} decrease its entropy by consuming energy from its environment.

Thus if we wish to speak of information flow and increase, this must be done relative to subsystems. Indeed, the fundamental objects of study should be \emph{open systems}, whose behaviour must be understood in relation to an external environment. Subsystems which can observe incoming information from their environment, and act to send information to their environment, have the capabilities of \emph{agents}.

\textbf{Moral:} Agents and their interactions are intrinsic to the study of information flow and increase in computation.
The classical theories of information do not  reflect this adequately.

\paragraph{Observer-dependence of information increase?}
Yorick Wilks (personal communication) has suggested the following additional twist. Consider an equation such as
\[ 3 \times 5 = 15 . \]
The forward direction $3 \times 5 \rightarrow 15$ is obviously a natural direction of computation, where we perform a multiplication. But the reverse direction $15 \rightarrow 3 \times 5$ is also of interest --- finding the prime factors of a number! So it seems that the \emph{direction of possible information increase} must  be understood as relative to the observer or user of the computation!

Can we in fact find an objective, observer-independent notion of information increase?
This seems important to the whole issue of whether information is inherently subjective, or whether it has an objective structure.

\subsection*{Problem 2}

The second puzzle is the computational version of what has been called the \emph{scandal of deduction} \cite{hintikka1970information,d2009enduring,sequoiah2008scandal}.
The logical problem is to find the sense in which logical deduction can be informative, since, by the nature of the process,  the conclusions are `logically contained' in the premises. So what has been added by the derivation? This is a rather basic question, which it is surprisingly difficult  to find a satisfactory answer to.

Computation can be modelled faithfully as deduction, whether in the sense of deducing the steps that a Turing maching takes starting from its initial configuration, or more directly via the Curry-Howard isomorphism \cite{Cur58,howard1980formulae}, under which computation can be viewed as performing cut-elimination on proofs, or normalization of proof terms. Thus the same question can be asked of computation: since the result of the computation is logically implied by the program together with the input data, what has been added by computing it?

The same issue can be formulated in terms of the logic programming paradigm, or of querying a relational database: in both cases, the result of the query is a logical consequence of the data- or knowledge-base.

It is, of course, tempting to answer in terms of the complexity of the inference process; but this seems to beg the question. We need to understand first what the inference process is doing for us!

We can also link this puzzle to another well-known issue in logic, namely the principle of \emph{logical omnisicience} in epistemic logic, which is unrealistic yet hard to avoid.
This principle can be formulated as follows:
\[ [K_a \phi \; \wedge \; (\phi \rightarrow \psi)] \; \rightarrow \; K_a \psi . \]
It says that the knowledge of agent $a$ is deductively closed: if $a$ knows a proposition $\phi$, then he knows all its logical consequences.
This is patently untrue in practice, and brings us directly back to our puzzle concerning computation.
We compute to gain information we did not have. We start from the information of knowing the program and its input, and the computation provides us with explicit knowledge of the output.
But what does `explicit' mean?

The computational perspective may indeed provide a usefully clarifying perspective on the issue of logical omniscience, since it provides a context in which the distinction between `explicit' and `implicit' knowledge can be made precise.
Let us start with the notion of a function.
In the 19th century, the idea of a function as a `rule' --- as given by some defining expression --- was replaced by its `set-theoretic semantics' as a set of ordered pairs. In other terminology, a particular defining expression is an \emph{intensional description} of a function, while the set of ordered pairs which it denotes is its \emph{extension}.

A program is exactly an intensional description of a function, with the additional property that this description can be used to explicitly calculate outputs from given inputs in a stepwise, mechanical fashion.\footnote{We refer e.g. to \cite{gandy1980church,sieg2002calculations} for attempts to give a precise mathematical characterization of `mechanical'.}
We can say that implicit knowledge, in the context of computation, is knowledge of an intensional description;
while explicit knowledge, of a data item such as a number, amounts to possessing the  \emph{numeral} (in some numbering system) corresponding to that number; or more generally, to possessing a particular form of intensional description which is essentially isomorphic to the extension. 


The purpose of computation in these terms is precisely to convert intensional descriptions into extensional ones, or implicit knowledge of an input-output pair into explicit knowledge.
The \emph{cost} of this process is calibrated in terms of the resources needed --- the number of computation steps, the workspace which may be needed to perform these steps, etc.
Thus we return to the usual, `common-sense' view of computation.
The point is that it rests on this distinction between intension and extension, or implicit vs. explicit knowledge.

Another important aspect of why we compute is \emph{data reduction}---getting rid of a lot of the information in the input.
Note that normal forms are in general \emph{unmanagably big} \cite{Vor97}.
Note also that it is \emph{deletion of data} which creates thermodynamic cost in computation \cite{Lan00}.
Thus we can say that much (or all?)  of the actual usefulness of computation lies in getting rid of the hay-stack, leaving only the needle.

The challenge here is to build a useful theory which provides convincing and helpful answers to these questions. In our view these puzzles, naive as they are, point to some natural questions which a truly comprehensive theory of computation, incorporating a  `dynamics of information', should be able to answer.

\section{What Do We Compute?}

The classical notion of computability as pioneered by Turing \cite{turing1937computable} focusses on the key issue of \emph{how} we compute; of what constitutes a computation. However, it relies on pre-existing notions from mathematics as to \emph{what} is computed: numbers, functions, sets, etc.

This idea also served computer science well for many years: it is perfectly natural in many situations to view a computational process in terms of computing an output from an input. This computation may be deterministic, non-deterministic, random, or even quantum, but essentially the same general paradigm applies.

However, as computation has evolved to embrace diverse forms and purposes: distributed, global, mobile, interactive, multi-media, embedded, autonomous, virtual, pervasive, \ldots
the adequacy of this view has become increasingly doubtful.

Traditionally, the \emph{dynamics} of computing systems --- their unfolding behaviour in space and time --- has been a mere means to the end of computing the function which specifies the algorithmic problem which the system is solving.\footnote{Insofar as the dynamics has been of interest, it has been in quantitative terms, counting the resources which the algorithmic process consumes --- leading of course to the notions of algorithmic complexity. Is it too fanciful to speculate that the lack of an adequate structural theory of processes has been an impediment to fundamental progress in complexity theory?} In much of contemporary computing, the situation is reversed: the \emph{purpose} of the computing system is to exhibit certain behaviour. The \emph{implementation} of this required behaviour will seek to reduce various aspects of the specification to the solution of standard algorithmic problems.

\begin{center}
\fbox{\textbf{What does the Internet compute?}}
\end{center}
Surely not a mathematical function \ldots

\subsection*{Why Does It Matter?}

We shall mention two basic issues in the theory of computation which become moot in the light of this issue.

There has been an enormous amount of work on the first, namely the theory of concurrent processes. Despite this huge literature, produced over the past four decades and more, no consensus has been achieved as to what processes \emph{are}, in terms of their essential mathematical structure. Instead, there has been a huge proliferation of different models, calculi, semantics, notions of equivalence.
To make the point, we may contrast the situation with the $\lambda$-calculus, the beautiful, fundamental calculus of functions introduced by Church at the very point of emergence of the notion of computability \cite{church1941calculi}. Although there are many variants, there is essentially a unique, core calculus which can be presented in a few lines, and which delineates the essential ideas of functional computation.
In extreme contrast, there are a huge number of process calculi, and none can be considered as definitive.

Is the notion of process too amorphous, too open to different interpretations and contexts of use, to admit a unified, fundamental theory? Or has the field not yet found its Turing?
See \cite{abramsky2006fundamental} for an extended discussion.

The second issue follows on from the first, although it has been much less studied to date.
This concerns the Church-Turing thesis of universality of the model of computation.
What does this mean when we move to a broader conception of what is computed?
And are there any compelling candidates? Is there a widely accepted universal model of interactive or concurrent computation?

As a corollary to the current state of our understanding of processes as described in the previous paragraphs, there is no such clear-cut notion of universality.
It is important to understand what is at issue here. 
If we are interested in the process of computation itself, the structure of interactive behaviour, then on what basis can we judge if one such process is faithfully simulated by another?
It is not of course that  there are no candidate notions of this kind which have been proposed in the literature; the problem, rather, is that there are far too many of them, reflecting different intuitions, and different operational and application scenarios.

Once again, we must ask: is this embarrassing multitude of diverse and competing notions a necessary reflection of the nature of this notion, or may we hope for an incisive contribution from some future Turing which will unify and organize the field?

\bibliographystyle{plain}

\bibliography{turbib}
\end{document}